\documentclass[doublespacing]{elsart}
\usepackage{natbib}
\usepackage{graphics}
\usepackage{graphicx}
\usepackage{epsfig}
\usepackage{amssymb}
\usepackage{lineno}

\begin{document}
\begin{frontmatter}

\title{Transport properties of graphene nanoribbon heterostructures}
\author{L. Rosales\thanksref{label1}\corauthref{cor1}},
\ead{luis.rosalesa@usm.cl}
\corauth[cor1]{Tel.:+56-32-2654434;
fax:+56-32-2797656}
\author{P. Orellana\thanksref{label2}},
\author{Z. Barticevic\thanksref{label1}},
\author{M.Pacheco\thanksref{label1}}

\address[label1]{Departamento de F\'{i}sica, Universidad T\'ecnica
F. Santa Maria, Casilla postal 110 V, Valpara\'iso, Chile}
\address[label2]{Departamento de F\'{i}sica, Universidad Cat\'olica
del Norte, Casilla 1280, Antofagasta, Chile}

\begin{abstract}
We study the electronic and transport properties of heterostructures
formed by armchair graphene nanoribbons with intersections of finite
length. We describe the system by a tight-binding model and
calculate the density of states and the conductance within the
Green's function formalism based on real-space renormalization
techniques. We show the apparition of interface states and bound
states in the continuum which present a strong dependence of the
heterostructure geometry. We investigate the effects on the
conductance of an external perturbation applied on the edges atoms
of the intersection region.
\end{abstract}


\end{frontmatter}

\section{Introduction}
\label{I} The fabrication under controlled experimental conditions
of graphene nanoribbons (GNRs) has attracted a lot of scientific
interest in the last decade\cite{Ohta}. GNRs are single atomic
layers which can be understood as an infinite unrolled carbon
nanotube. The special electronic behavior of GNRs defined by their
quasi one-dimensional electronic confinement and the shape of the
ribbons ends (two cases of maximum symmetries can be obtained:
zigzag and armchair ends\cite{Nakada}), suggest remarkable
applications in graphene-based devices\cite{Shi}. Moreover, due to
the flat structure GNRs seem to be easier to manipulate than carbon
nanotubes. The transport properties of these structures have been
studied with special interest by several scientific groups. Peres et
al\cite{Peres} find that for clean systems, the quantization
condition for the electronic conductance is different for zigzag and
armchair GNRs . Shi et al\cite{Shi} show that applying a gate
voltage to the ribbons it is possible to obtain an electronic switch
very useful to study the Klein paradox.

In this work we present a theoretical description of the electronic
and transport properties of graphene heterostructures formed by
armchair GNRs of different widths. The systems are composed by M
unit cells of a semiconducting $N_{C}$ armchair GNR  surrounded by
two semi-infinite $N_{L}$ metallic ribbons  [with the condition
$N_{C}$ = $N_{L}$+4]. These kinds of systems allow us to consider
two possible configurations: a Cross-bar and a T-shaped
heterostructure, which are schematically shown in Fig.1\cite{Shi}.
We investigate the effects of a weak external perturbation applied
on the edges atoms of these two heterostructures. The external
perturbation slightly modifies the on-site energies of these atoms,
allowing the mixture of continuous and localized states of the
system. We found the presence of two kind of localized states:
interface states and bound states in the continuum. The former
states reside in the interface of the two perpendicular nanoribbons.
The latter states are bound states due to the quantum interference
of the electron wavefunction inside of the nanoribbons
intersection\cite{nockel,triple}. The external potential allows the
contribution of these localized states to the transmission of the
electrons through the central conductor which can be observed in the
conductance of these systems.

\begin{figure}[h!]
\begin{center}
\includegraphics [width=6.6 cm, height=7.5 cm]{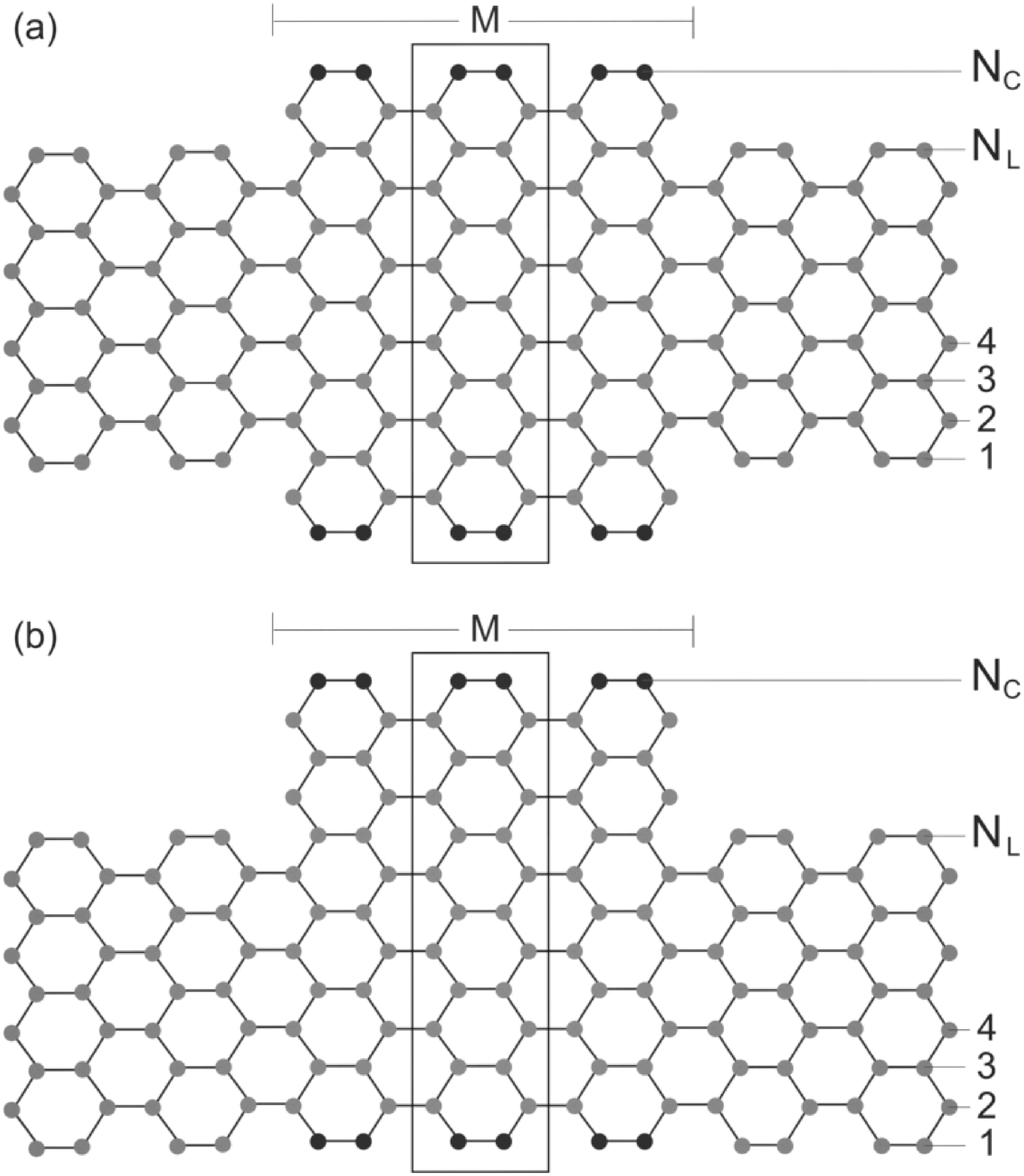}
\end{center}
\caption{Atomic configuration of the armchair heterostructures: (a)
Cross-bar and (b) T-shaped configuration. The black atoms at the
edges of the central region represent the effect of the applied
external perturbation $\Delta V$. The black rectangle define the
unit cell of the central part.}
\end{figure}

\section{Model}
\label{II} The systems are described in a real-space picture which
allows to incorporate the potential fluctuations at the microscopic
scale. We adopt a single $\pi$-band tight-binding Hamiltonian taking
into account only nearest-neighbor hopping interaction (with
$\gamma\approx 2.75 eV$). The density of states (DOS) and the
conductance of the systems are obtained in the Green's function
formalism based on real-space renormalization
techniques\cite{rocha}. The conductance is calculated using the
surface Green's functions matching formalism\cite{Nardelli}. Within
this picture the full system is partitioned into three parts: the
central structure (the GNR intersection region) and two leads (two
semi-infinite metallic GNRs). In the linear response approach, the
conductance can be calculated using the Landauer formula given by:
$\Gamma (E_F)= \frac{2e^2}{h} T(E_F)$, where $T(E_F)$ is the
transmission function of an electron crossing through a central
conductor (for more details see ref.\cite{Nardelli,datta,rocha}).
The strength of the external perturbation is written in term of the
potential difference between the edges atoms of the central
structure, $\Delta V$. In what follows all the energies are written
in terms of the hopping parameter $\gamma$ and the Fermi energy
level is taken as the zero of the energies.

\section{Results and Discussion}
\label{IV} Results of the DOS for the heterostructures formed by M=3
unit cells of $N_{C}$=9 and leads with $N_{L}$=5, for different
external potential strengths $\Delta V$, are shown in Fig. 2. In the
lower panel we show the behavior of the DOS of the two considered
structures as a function of the energy. We include in the plot the
DOS of a pristine armchair GNRs N=5 (solid black online) for
comparison. For the cross-bar heterostructure (dash blue online),
two kind of states can be distinguished: interface states and bound
states in the continuum (resonant states). The interface states come
from the junction region between the leads and the central ribbon
structure. These states are pinned at the Van Hove singularities of
the pristine N=5 armchair GNR . On the other hand, the bound states
in the continuum (two symmetric states with respect to the Fermi
level at $\pm 0.368 \gamma$) arise from the interference of the
electron wave function inside the central ribbon. The occurrence of
these states is only an effect of the space symmetry of the
heterostructure configuration. This can be corroborated comparing
the behavior of the DOS of the T-shaped structure (dot red online)
with the cross-bar structure. It is possible to observe the
interface states at the same energies as that for the cross-bar
heterostructure but the bound states disappear due to the breaking
of the spatial symmetry of the system \cite{zhen-li}. We analyze the
effect of the external potential strength $\Delta V$ on the bound
states and interface states of the heterostructures. In the central
panel we show the behavior of one bound state ($0.368\gamma$) and
one interface state ($1.0\gamma$) of the cross-bar heterostructure
for different potential strengths $\Delta V$. It is possible to
observe that the bound state is not degenerated and the effect of
the external potential is only a slight displacement of the energy
level proportional to $\Delta V$. On the other hand, besides of the
energy shift, the interface state shows a break of its degeneracy at
$\Delta V$=0.06$\gamma$. This fact affects the conductance of the
system as we see later. In the upper panel we show the behavior of
the interface state pinned at $1.0\gamma$ for the T-shaped
heterostructure. As expected due to its nature, the same behavior of
the cross-bar heterostructure is obtained, a break of the degeneracy
and an energy shift proportional to $\Delta V$ can be observed.

\begin{figure}[h!]
\begin{center}
\includegraphics {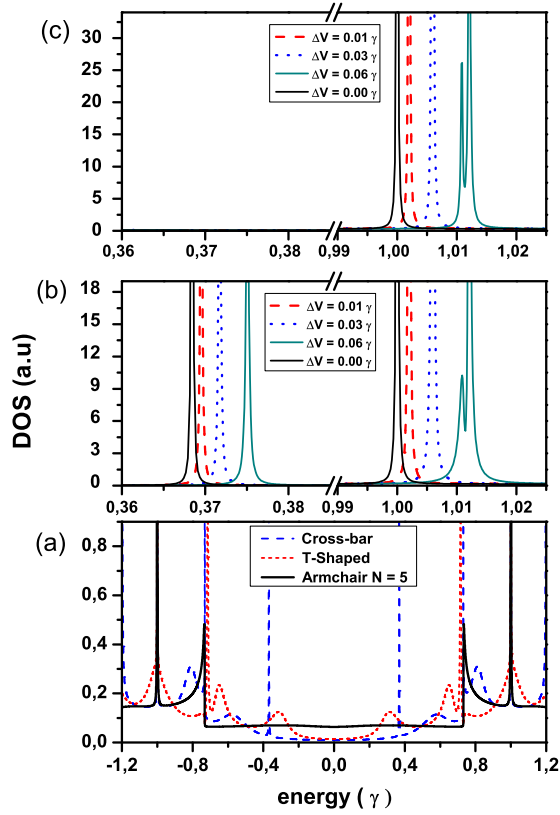}
\end{center}
\caption{(a) DOS as a function of the energy of the cross-bar and
the T-shaped heterostructures in comparison with the DOS of a
pristine armchair GNR N=5, (b) DOS of the cross-bar heterostructure
for two special energy values for different external potential
strength and (c) DOS of the T-shaped heterostructure for a special
energy values for different external potential strength $\Delta V$.}
\end{figure}

In what follow we consider the behavior of the conductance of the
heterostructures described before as a function of the external
perturbation potential. Results of the conductance for the cross-bar
heterostructures $N_{L}=5$ and $3\times$$N_{C}=9$, for different
external potential strengths $\Delta V$ are shown in Fig. 3. For all
considered values of $\Delta V$ the system shows a semiconductor
behavior\cite{Shi}. A Fano antiresonance pinned at $0.47\gamma$ and
a thin Fano resonance pinned at $0.65\gamma$, which arise from the
electron interference inside of the central structure, are almost
invariant when the external potential is applied over the cental
structure. For $\Delta V$ $>$ 0 very thin peaks and dips can be
observed in the conductance curves at defined energy levels. For
instance, at energy equal to $1.0\gamma$ the conductance shows an
evolution between peaks and dips depending on the strength of
$\Delta V$. This behavior is due to the shift and break of
degeneracy of the interface states at that energy. The intensity of
the dips becomes a quantum of conductance for $\Delta V$ $\geq$
$0.06\gamma$. For energy equal to $0.65\gamma$ a thin peak in the
conductance is observed, which evolves with the potential strength.
These peaks are produced by a bound state that becomes a resonant
one for finite $\Delta V$ values and it starts to contribute to the
transmission of the system when the external potential is applied.

\begin{figure}[h!]
\begin{center}
\includegraphics {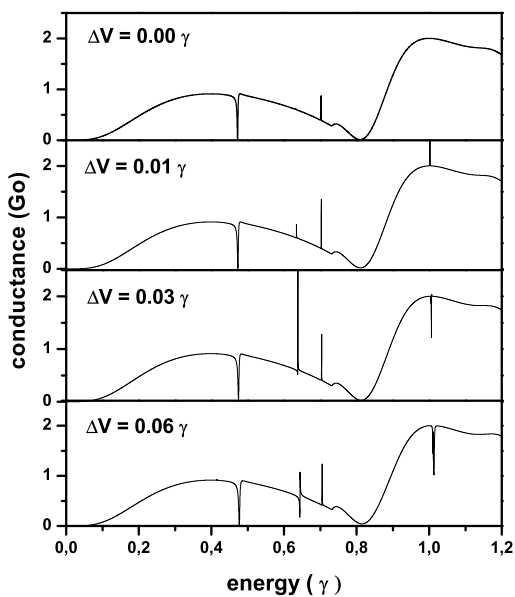}
\end{center}
\caption{Evolution of the conductance of the cross-bar
heterostructure for different values of the external potential
strength}
\end{figure}

Results of the conductance for the T-shaped heterostructures for
different potential strengths $\Delta V$ are shown in Fig. 4.
Similar to the cross-bar heterostructures, the system has a
semiconductor behavior for all considered $\Delta V$ values. In
comparison with the cross-bar heterostructure, the conductance shows
more antiresonances (i.e point of zero conductance). These
antiresonances arise from the Fano effect due to the destructive
interference of the electron when it follows different pathways. It
is possible to observe at energy equal to $1.0\gamma$, a thin peak
that evolves with the external perturbation, which grows in
intensity and becomes narrower due to the break of the degeneracy of
the interface state pinned at this energy level when the potential
strength increase.

\begin{figure}[h!]
\begin{center}
\includegraphics {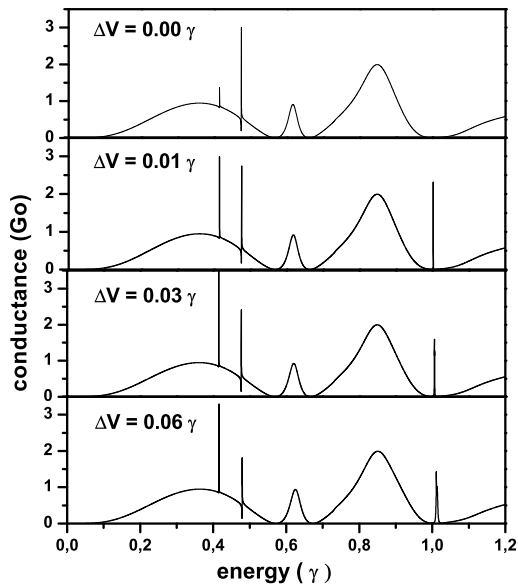}
\end{center}
\caption{Evolution of the conductance of the T-shaped
heterostructures for different values of the external potential
strength}
\end{figure}

\section{Summary}
\label{IV} We have calculated the DOS and conductance of
heterostructures formed by armchair GNRs. We analyzed their
electronic properties when a weak external potential is applied on
the edges atoms of the central ribbon structure. We found the
presence of interface states and bound states in the continuum. This
corresponds to the formation of collective states by the coupling of
individual states and the interference among the intersection
region.

\section{Acknowledgments}
\label{III} This work was supported by CONICYT/Programa Bicentenario
de Ciencia y Tecnologia (CENAVA, grant ACT27).


\begin{thebibliography}{10}

\bibitem{Ohta} T. Ohta et al,Science \textbf{313} (2006) 951 ;
S. Stancovich et al, Nature \textbf{442} (2006) 282.

\bibitem{Nakada} K. Nakada et al, Phys. Rev B \textbf{54} (1996) 17954 ; K. Wakabayashi
and H. Hiroshima, Phys. Rev B \textbf{64} (2001) 125428.

\bibitem{Shi} Q. W. Shi et al, Cond-Mat/0611604 v1, (2006); A. Rycerz et
al, cond-mat/06080533 v1, (2006).

\bibitem{Peres} N. Peres, A. Castro and F.Guinea Phys. Rev B \textbf{73} (2006)
195411.

\bibitem{nockel} J. U. N\"{o}ckel, Phys. Rev. B \textbf{46} (1992) 15348.

\bibitem{zhen-li} Zhen-Li Ji and Karl-Frederik Berggren, Phys.
Rev. B \textbf{45} (1992) 6652.

\bibitem{triple} M. L. Ladr\'on de Guevara and P. Orellana, Phys. Rev. A \textbf{73} (2006)
205303.

\bibitem{rocha} C. G. Rocha, A. Latg\'e, and L. Chico, Phys. Rev. B \textbf{72} (2005) 085419
; L. Rosales, C. G. Rocha, A. Latg\'e, Z. Barticevic and M. Pacheco,
Phys. Rev. B\textbf{75} (2007) 165401.

\bibitem{Nardelli} M. Nardelli, Phys. Rev B \textbf{60} (1999) 7828.

\bibitem{datta} S. Datta, \emph{Electronic Transport in Mesoscopic Systems}, Cambridge
University Press, 1995.

\end{thebibliography}
\end{document}